# Strain-induced Evolution of Electronic Band Structures in a Twisted Graphene Bilayer

Wei Yan[1,§], Wen-Yu He[1,§], Zhao-Dong Chu[1,§], Mengxi Liu[2], Lan Meng[1], Rui-Fen Dou[1], Yanfeng Zhang[2,3], Zhongfan Liu[2], Jia-Cai Nie[1], and Lin He[1,*]

Here we study the evolution of local electronic properties of a twisted graphene bilayer induced by a strain and a high curvature. The strain and curvature strongly affect the local band structures of the twisted graphene bilayer; the energy difference of the two low-energy van Hove singularities decreases with increasing the lattice deformations and the states condensed into well-defined pseudo-Landau levels, which mimic the quantization of massive Dirac fermions in a magnetic field of about 100 T, along a graphene wrinkle. The joint effect of strain and out-of-plane distortion in the graphene wrinkle also results in a valley polarization with a significant gap, i.e., the eight-fold degenerate Landau level at the charge neutrality point is splitted into two four-fold degenerate quartets polarized on each layer. These results suggest that strained graphene bilayer could be an ideal platform to realize the high-temperature zero-field quantum valley Hall effect.

[1] Department of Physics, Beijing Normal University, Beijing, 100875, People's Republic of China. [2] Center for Nanochemistry (CNC), College of Chemistry and Molecular Engineering, Peking University, Beijing 100871, People's Republic of China. [3] Department of Materials Science and Engineering, College of Engineering, Peking University, Beijing 100871, People's Republic of China.
[§]These authors contributed equally to this paper.
Correspondence and requests for materials should be addressed to L. H. (email: helin @bnu.edu.cn)

Graphene, a material made of a two-dimensional (2D) honeycomb carbon lattice, is acting as a bridge between quantum field theory and condensed matter physics due to its gapless, massless, and chiral Dirac spectrum[1-8]. Graphene's 2D nature makes it amenable to external mechanical deformation. It is well-established by now that lattice deformations of graphene lead to the appearance of a pseudomagnetic gauge field acting on the charge carriers[4,9-16]. The pseudomagnetic field becomes an experimental reality after the observation of Landau levels (LLs) in a strained graphene monolayer[17,18] and in a strained artificial graphene[19]. The energies of these LLs follow the progression of LLs of massless Dirac Fermions, which suggest a possible route to realize zero-field quantum-Hall effects in the strained graphene[17-19]. Compared with the single-layer graphene, the graphene bilayers, including AA stacked, AB stacked (Bernal), and twisted graphene bilayer, display even more complex electronic band structures and intriguing properties because of the interplay of quasiparticles between the Dirac cones on each layer[20-33]. Recently, several groups addressed the physics of the strained graphene bilayer (either AA or AB stacked graphene bilayer) theoretically and obtained many interesting results[34-40]. Despite many suggestive findings and potential applications, there have unfortunately been no experimental studies of the effect of strain on the electronic band structures of the graphene bilayer.

In this article, we report the local electronic properties of a corrugated twisted graphene bilayer with a twist angle $\theta \sim 5.1°$ studied by scanning tunneling microscopy and spectroscopy (STM and STS). In the flat region of the sample, we observed two low-energy van Hove singularities (VHSs), which originate from two saddle points flanking the Dirac points of the band structure[24,26,27,31,32], and superlattice Dirac points generated by the graphene-on-graphene moiré. Around a wrinkle of the twisted graphene bilayer, the moderate lattice deformations and the possible enhanced interlayer coupling lead to the decrease of the energy difference of the two VHSs. Remarkably, we observed apparent zero-field Landau level-like quantizations, which mimic the quantizations of massive Dirac Fermions of Bernal graphene bilayer in a perpendicular magnetic field of about 100 T, along the strained wrinkle. We also show that the strain and curvature result in a valley polarization of the twisted graphene bilayer, i.e., the eight-fold degenerate LL at the charge neutrality point is splitted into two four-fold degenerate quartets polarized on each layer with a significant gap. Our experimental result suggests that the strained twisted graphene bilayer could be an ideal platform to realize zero-field quantum valley Hall effect[41,42].

**Results**

**Strained twisted graphene bilayer around a Rh step edge.** The twisted graphene bilayer was grown on a 25 micron thin Rh foil via a traditional ambient pressure chemical vapor deposition (CVD) method[31], and only the sample mainly covered with two graphene layers was further studied by STM and STS. Due to the thermal expansion mismatch between graphene and the substrate, defect-like wrinkles and ripples tend to evolve along the boundaries of crystalline terraces for a strain relief[17,43,44]. Briefly, the sample was synthesized at 1000 ºC and after growth when the sample is cooled down, the graphene expands while the Rh foil contracts. As a consequence, strain builds up in the sample and the graphene wrinkles are induced at the step edges of Rh foil where the graphene is possibly weakly coupled to the Rh.

Figure 1a shows a typical STM image of a graphene wrinkle at the boundary of two flat terraces of the Rh surface. Moiré patterns with identical period ~ 2.76 nm are observed on both the right and left terraces flanking the wrinkle, as shown in Fig. 1b. For monolayer graphene on a Rh(111) surface, the lattice mismatch between graphene (0.246 nm) and Rh(111) (0.269 nm) leads to hexagonal moiré superstructures with the expected periodicity ~ about 2.9 nm resulted from a 12C/11Rh coincidence lattice[45-47] (see Supplementary Information, the characteristic of the moiré superstructures is distinct from that shown in Fig. 1.). The 2.76 nm periodic protuberance, as shown in Fig. 1, is attributed to the moiré pattern arising from a stacking misorientation between the top graphene layer and the underlayer graphene. The twisted angle $\theta$, estimated as 5.1°, is related to the period of the moiré pattern by $D = a/[2\sin(\theta/2)]$. The line profile, as shown in Fig. 1c, indicates that the one-dimensional wrinkle is a strained structure of the twisted graphene bilayer with a high curvature and out-of-plane distortions at the apex (see Supplementary Information for a STM topography of the wrinkle). The average height and width (peak width at half-height) of the wrinkle are about 7.9 nm and 8.2 nm respectively. For a corrugated twisted graphene bilayer with a small curvature, we can observe moiré superstructures along wrinkles and ripples[43,44] (see Supplementary Information). However, it is difficult to obtain atomic-resolution image of the wrinkle shown in Fig. 1 because of its large curvature. The wrinkle could still be twisted graphene bilayer with a finite interlayer coupling and, of course, the large bending of the wrinkle also can result in a local slide of two graphene sheets relative to each other, then, the two graphene sheets may become decoupled at the very top of the wrinkle.

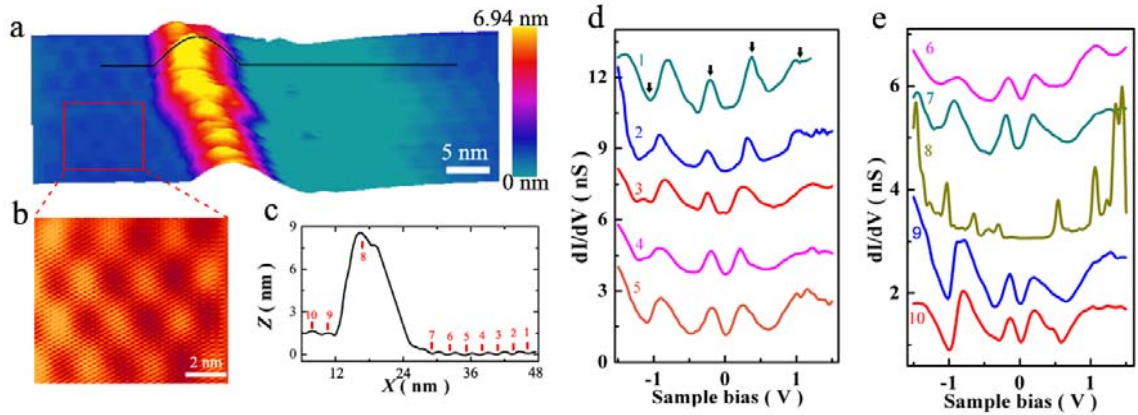

**Figure 1 | STM images and STS of a twisted graphene bilayer with a strained wrinkle along a step of the Rh foil. a,** Large-area STM image of a twisted graphene bilayer with a wrinkle along a step of the Rh foil ($V_{sample}$ = -600 mV and I= 0.24 nA). Moiré pattern with identical period appears on both the right and left terraces flanking the wrinkle. **b,** Atomic-resolution image of the twisted graphene bilayer in the red frame of panel (**a**). It shows a moiré pattern with the period of 2.76 nm ($V_{sample}$ = -351 mV and I = 0.14 nA). The twisted angle of the graphene bilayer is estimated as about 5.1°. **c,** A line profile along the black curve in panel (**a**). The average height and width (peak width at half-height) of the wrinkle are about 7.9 nm and 8.2 nm respectively. The small periodic protuberances with a period of about 2.76 nm on the two terraces are attributed to the moiré pattern. STS measurements at different positions along the line profile show quite different characteristics, as shown in (**d**) and (**e**). The spectra have been vertically offset for clarity. Curves 1, 2, 3, 4, 5, 6, 7, 9, and 10 are measured on the flat twisted graphene bilayer. The two peaks flanking the Dirac point, as pointed out by the arrows, are attributed to the two VHSs. The two dips pointed out by the arrows are the positions of the superlattice Dirac points generated by the graphene-on-graphene moiré. The sharp peaks in curve 8 are attributed to the Landau quantizations of the strained graphene bilayer in a large pseudomagnetic field.

**VHSs and superlattice Dirac points of the strained twisted graphene bilayer.** Figure 1d and 1e show ten dI/dV-V curves recorded at different positions along the line profile in Fig. 1c. Curves 1-7, 9, and 10 are measured at the flat twisted graphene bilayer on the Rh terraces. Curve 8, which shows distinct characteristics comparing to that of the curves 1-7, 9, and 10, is measured at the strained wrinkle. The tunneling spectrum gives direct access to the local density of states (LDOS) of the surface at the position of the STM tip. The experimental result in Fig. 1d and 1e indicates that the strain affects the electronic band structures of the twisted graphene bilayer remarkably.

For a twisted graphene bilayer, the Dirac points of the two layers no longer coincide and the zero energy states occur at k = -$\Delta K$/2 in layer 1 and k = $\Delta K$/2 in layer 2 (here $\Delta K$ = $2K\sin(\theta/2)$ is the shift between the corresponding Dirac points of the twisted graphene bilayer, and $K = 4\pi/3a$ with a ~ 0.246 nm the lattice constant of the hexagonal lattice).

When there is a finite interlayer hopping $t_\theta$, two saddle points (two VHSs) at $\Delta E_{vhs} = \hbar v_F \Delta K - 2t_\theta$ are unavoidable along the intersection of the two Dirac cones: $K$ and $K_\theta$[26, 27,31,48,49] (see Supplementary Information, here $v_F$ is the Fermi velocity). The two pronounced peaks flanking zero-bias in the tunneling spectra of the curves 1-7, 9, and 10, as shown in Fig. 1, are attributed to the two VHSs of the LDOS. At a fixed position, the energy difference of the two VHSs $\Delta E_{vhs}$ deduced from the tunneling spectra is almost a constant. However, the energy difference of the two peaks decreases when the experimental position approaches the strained wrinkle (similar experimental result was also observed in the left terraces of the wrinkle). For the curve 1, the energy difference of the two VHSs is $\Delta E_{vhs} \sim 0.59$ eV. For the curve 7, the value of $\Delta E_{vhs}$ decreases to 0.38 eV. We will demonstrate subsequently that the decrease of $\Delta E_{vhs}$ originates from the strain-induced lattice deformations.

Besides the two VHSs, the curves 1-7, 9, and 10, show two dips in the LDOS (marked by arrows), symmetrically placed at about $\pm (1.05\pm0.05)$ eV around the graphene Dirac point, irrespective of the experimental positions, but generally of asymmetric strength. This reminds us the characteristic of superlattice Dirac points at $E_{SD} \sim \pm \hbar v_F |G|/2$ in both graphene monolayer and twisted graphene bilayer induced by a weak periodic potential (experimentally, the periodic potential can be generated by the moiré pattern between the top layer graphene and the substrate (or the under layer graphene); here G is the reciprocal superlattice vectors of the moiré pattern)[50,51]. The graphene-on-graphene moiré pattern can provide a weak periodic potential, which leads to the emergence of the superlattice Dirac points[51,52]. Considering the period of the moiré pattern ~ 2.76 nm and $v_F$ ~ $1.1 \times 10^6$ m/s, we obtain $E_{SD}$ ~ ± 0.95 eV, which is slightly smaller than our experimental value (see Supplementary Information for the theoretical model of the emergence of superlattice Dirac points in the twisted graphene bilayer in a periodic potential). This slight discrepancy may arise from the fact that the value of velocity far from the charge neutrality point of graphene is larger than $1.1 \times 10^6$ m/s. It could reach $1.5 \pm 0.2 \times 10^6$ m/s[53].

**Landau level quantizations and valley polarization in the graphene wrinkle**. Now we turn to understand the distinct spectrum obtained at the strained wrinkle, shown as the curve 8 in Fig. 1. The sharp peaks of the tunneling spectrum are attributed to the LLs of the strained graphene wrinkle. The large shear stress, strain stress, and out-of-plane distortions of the graphene wrinkle can result in a large pseudomagnetic field on its electronic structures[54-57]. Our analysis is summarized in Fig. 2. It is shown that these LLs follow the progression of LLs of the massive Dirac fermions $E_N = \pm \{\hbar\omega_c[N(N-1)]^{1/2} + E_g/2\}$, N = 0,1,2,…. (Here $\omega_c = eB_S/m^*$ is the cyclotron frequency, $B_S$ is the pseudomagnetic field, $m^* = 0.03m_e$ with $m_e$ the mass of electron, and $E_g$ ~ 0.32 eV is the band gap)[20,30]. The energy of the pseudo-Landau levels as a function of the orbital and valley index (N, ξ) are plotted in Fig. 2(d). It shows that our experimental data match the equation perfectly. The pseudomagnetic field of the strained wrinkle estimated according to the spectrum is about

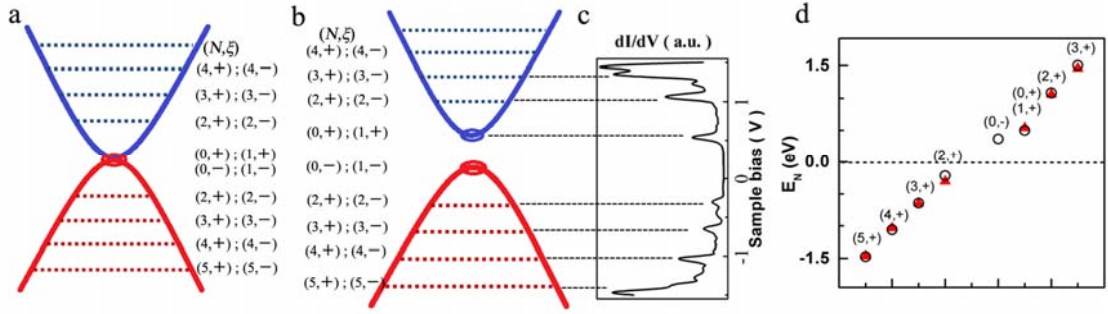

**Figure 2| Landau quantizations of the massive Dirac fermions.** The formation of LLs of massive Dirac fermions in the quantum Hall regime (**a**) without and (**b**) with a bandgap. LLs are indexed by the orbital and valley index, (N,ξ) [ξ =+1 for valley **K**, ξ =-1 for valley **K'**]. When there is a gap, the eight-fold degenerate level at the charge neutrality point becomes layer-polarized quartets. The $LL_{(0,+);(1,+)}$ projected on the top layer and the $LL_{(0,-);(1,-)}$ on the underlayergraphene. (c) The tunneling spectrum at the strained wrinkle. The sharp peaks are attributed to the LLs. The energies of these LLs follow the progression of LLs of the massive Dirac Fermions with a band gap. The pseudomagnetic field is estimated as 100 T and the gap of the valley polarization is about 0.32 eV. （d）The pseudo-Landau levels (the solid triangles) of the experimental data in (c) as a function of the orbital and valley index (N, ξ). The open circles show the progression of LLs of massive Dirac Fermions $E_N = \pm \{\hbar\omega_c[N(N-1)]^{1/2} + E_g/2\}$, N = 0,1,2,......

100 T (see Supplementary Information). The wrinkle is one of the simplest model to study the strain-induced pseudo-magnetic flux of a corrugated graphene sheet[13]. The flux of the wrinkle can be estimated by $\Phi = (\beta h^2/la)\Phi_0$ (here $h$ is the height, $l$ is the width of the wrinkle, $2 < \beta < 3$, $a$ is on the order of the C-C bond length, and $\Phi_0$ is the flux quantum)[13,17,18]. It is unexpected that the strain-induced Landau level-like quantizations of the twisted graphene bilayer are identical to the Landau quantizations of the massive Dirac Fermions in Bernal stacking graphene bilayer in an external magnetic field[20]. This result suggests that the corrugated structure modifies structure of the twisted graphene bilayer along the wrinkle efficiently, which as a consequence results in the massive chiral fermions (where the electronic energy dispersion is hyperbolic in momentum) of the strained wrinkle. We will demonstrate subsequently that the strain and the large curvature could lead to a hyperbolic band structure of the graphene wrinkle.

The size of the LL wavefunction, i.e., the magnetic length $l_B = [\hbar/(eB)]^{1/2}$, generated by the pseudomagnetic field (~ 100 T) is about 2.6 nm, which is much smaller than the size of the graphene wrinkle. The difference between on-site potential (the staggered potential) of the A and B atoms in the two graphene sheets becomes important along the graphene wrinkle[29]. Therefore, it is expected that the eight-fold degenerate LL at the charge neutrality point could be locally lifted by the pseudomagnetic field. In our experiment, the eight-fold degenerate LL at the charge neutrality point is splitted into two four-fold degenerate quartets polarized on each layer with a gap $E_g$ ~ 0.32 eV. Importantly, the $N = 0$ and $N = 1$ states $LL_{(0,+);(1,+)}$ in the ***K*** valley (labeled as $\xi = +1$) are localized predominantly on the B sites of the top layer, whereas the $N = 0$ and $N = 1$ states $LL_{(0,-);(1,-)}$ in the ***K'*** valley (labeled as $\xi = -1$) are localized predominantly on the A sites of the underlayergraphene[29]. Therefore, it is not surprised that only $LL_{(0,+);(1,+)}$ in the ***K*** valley, which is polarized on the top layer, could be observed by the STS measurement, as shown in Fig. 2 (see Supplementary Information). The energy interval with zero tunneling DOS around the Fermi level of the spectrum is therefore the sum of the energy gap and the energy difference between the $LL_{(0,-);(1,-)}$ and $LL_{(2,+);(2,-)}$. Very recently, it is predicted that the valley polarized quantum Hall phases can be driven by Coulomb interaction in strained graphene[58]. The observed valley polarization (with the energy gap ~ 0.32 eV) on each layer and large pseudomagnetic field suggest that the strained twisted graphene bilayer is an ideal platform to realize high-temperature zero-field quantum valley Hall effect.

**Discussions**

To further understand our experimental result, we calculated the effects of both the strain and curvature on the electronic band structures of the twisted graphene bilayer by tight binding model with the Hamiltonian $H = H_1 + H_2 + H_\perp$, where $H_1$ and $H_2$ are the Hamiltonians for each layer and $H_\perp$ is the interaction Hamiltonian between the two layers[24,48,59]. In the flat region around the wrinkle, the lattices of the twisted graphene bilayer are compressed several percents in a prescribed direction and the out-of-plane distortion of the sample is very small. Therefore, we only consider the effect of a tensional strain on the electronic structure of the twisted graphene bilayer. Figure 3a-3c shows the energy dispersions of the twisted graphene bilayer with different lattice deformations. With increasing lattice deformations and the variation of the nearest-neighbor hopping parameters, the Dirac points move away from the corners *K* and *K'*, as shown in Fig. 3e. Figure 4c summarizes the value of $\Delta E_{vhs}$ of the flat twisted graphene bilayer as a function of the tensional strain. In small and moderate deformations, the energy difference of the two VHSs $\Delta E_{vhs}$ decreases as the strain increases, which agrees well with our experimental result.

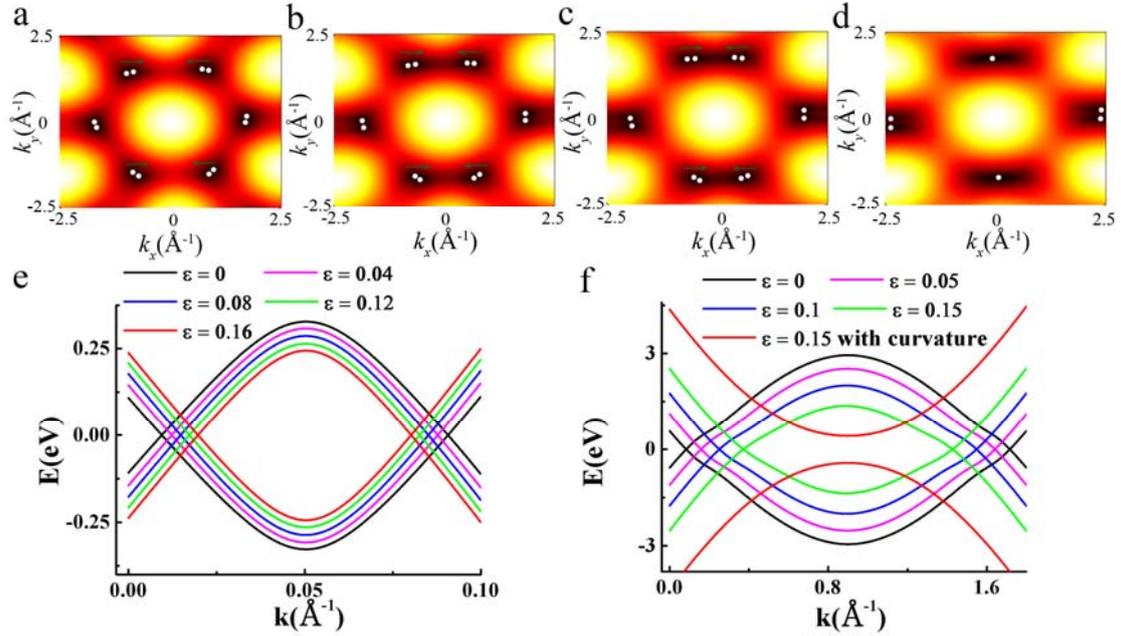

**Figure 3 | Electronic band structures of a twisted graphene bilayer.** Density plots of the energy dispersion of twisted graphene bilayer for (**a**) {$\varepsilon = 0, \theta = 5.1°$}, (**b**) {$\varepsilon = 0.1, \theta = 5.1°$}, (**c**) {$\varepsilon = 0.15, \theta = 5.1°$}, (**d**) {$\varepsilon = 0.15, \theta = 5.1°$} with a curvature radius ~ 1.41 nm at the top of the wrinkle. The white points are the four Dirac points: $K$, $K_\theta$, $K'$, and $K'_\theta$. In panel (**e**) we have a cut of the electronic band structures along $K$ and $K_\theta$, showing the decreasing of $\Delta E_{vhs}$ as strain increases (in a small and moderate strain). In panel (**f**) we have a cut of the electronic band structures along $K$ and $K_\theta'$, showing the merging of the Dirac cones as the strain increases and the ultimate appearance of the gap induced by the joint effect of the strain and the curvature.

On the top of the wrinkle, the large bending causes rehybridization between π and σ orbitals of graphene and influences the Slater-Koster overlap integral[3,12,54,60,61]. Therefore, both the effect of the strain and the curvature should be considered in the calculation (see Supplementary Information for details). We fit the section line of the wrinkle by a Gaussian curve to estimate its curvature, as shown in Fig. 4a. The large curvature at the top of the wrinkle could merge the Dirac points *K* and *K'* into a single one (Figure 3d). The joint effects of a considerable lattice deformation and the large curvature result in an almost hyperbolic electronic band of the wrinkle (Figure 3f). As a consequence, the Landau level-like quantizations in the graphene wrinkle should be identical to the Landau quantizations of the massive Dirac Fermions in Bernal stacking graphene bilayer (as shown in Fig. 2). A band gap also appears in the energy spectrum of the graphene wrinkle, as shown in Fig. 3f. This provides a physical mechanism to open a significant energy gap in the twisted graphene bilayer where an external electric field can not open a gap in its spectrum[24]. Figure 4b shows the energy gap $E_g$ as a function of the strain for a fixed curvature radius ~ 1.41 nm. The large curvature at the top of the wrinkle opens an energy gap in twisted graphene bilayer when the amount of strain ε reaches about 0.14.

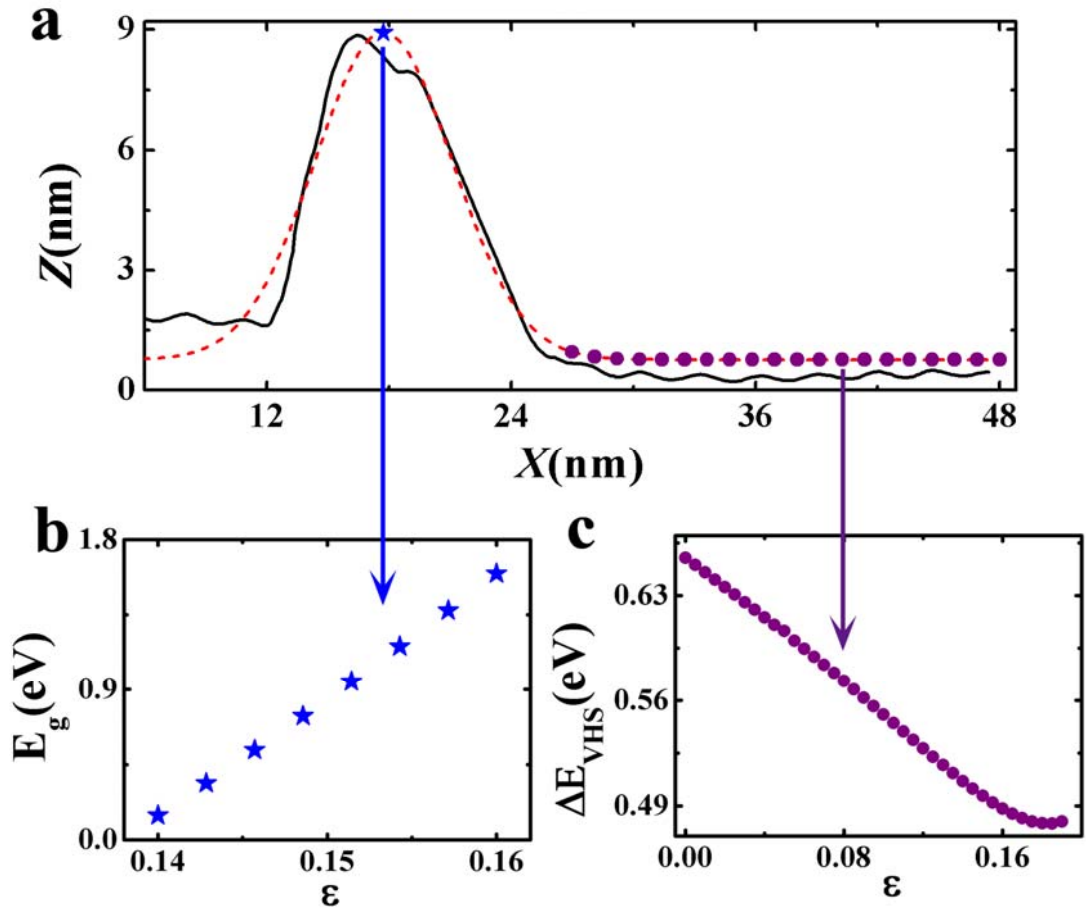

**Figure 4 | The energy difference of the two VHSs $\Delta E_{vhs}$ and the energy gap as a function of strain.** In panel (**a**), the red dash line corresponds to a Gaussian fitting to the section line of the graphene wrinkle, as shown in Figure 1 (c). Panel (**b**) shows the energy gap increases as the strain increases for a fixed curvature radius ~ 1.41 nm at the top of the wrinkle. Panel (**c**) shows the value of $\Delta E_{vhs}$ as a function of the strain in flat twisted graphene bilayer.

The pseudo-Landau quantizations observed in our experiment may be induced by the local nonuniform strains since that the strained structure is quite rough along the wrinkle (see Supplementary Information). Usually, a constant pseudomagnetic field can be achieved when the honeycomb lattice are transversely displaced from their original positions in the following way $(u_r, u_\phi) = qr^2(\sin3\phi, \cos3\phi)$, where $u_r$ and $u_\phi$ are the radial and azimuthal displacements, $\phi$ is the azimuthal angle, $r$ is the distance from anbitrary origin, and $q$ is a parameter corresponding to the strength of the strain[14,17,19]. This uniform pseudomagnetic field is not easy to realize in the strained graphene[17,18]. Our experiment also confirms that the pseudomagnetic field is not uniform along the wrinkle (see Supplementary Information). However, the nonuniform strains can generate large local pseudomagnetic fields, which can be detected by the STS measurements along the wrinkle.

In summary, we studied the local electronic properties of the strained graphene bilayer with a twist angle ~ 5.1°. Our experiment demonstrated that the energy difference of the two VHSs $\Delta E_{vhs}$ decreases as the strain increases. Along the wrinkle of the twisted bilayer graphene, we observed apparent zero-field Landau level-like quantizations and the valley polarization with a large gap. The observed valley polarization on each layer and large pseudomagnetic field ~ 100 T suggest that the strained graphene bilayer is an ideal platform to realize high-temperature zero-field quantum valley Hall effect. Further effort should be made to generate large-scale strained graphene bilayer with a uniform pseudomagnetic field to carry out the transport measurements in this interesting system.

**Methods**

**Preparation of bilayer graphene on Rh foil.** The graphene bilayer was grown on a 25 micron thin Rh foil via a traditional ambient pressure chemical vapor deposition (CVD)method[31]. The process is similar to the systhesis of graphene on Pt and Cu foils, which were reported in previous papers[43,44]. Briefly, the Rh foil was firstly heated from room temperature to 1000ºC in 45 min under an Ar flow of 850 sccm. Then the furnace was experienced a hydrogen gas flow of 50 sccm for 40 min at 1000 ºC. Finally, $CH_4$ gas was introduced with a flow ratio of 5-10 sccm, and the growth time is varied from 3 to 15 min for controlling the thickness of graphene. The as-grown sample is cooled down to room temperature andcan be transferred into the ultrahigh vacuum condition for further characterizations. The thickness of the as-grown graphene was characterized by Raman spectra measurements[31], and only the sample mainly covered with graphene bilayer was further studied by STM and STS. The Raman spectroscopy for the sample used here is shown in Fig. S1 of Ref. 31. In previous papers, we studied carefully the structures and the formation mechanism of the wrinkles and ripples on metallic substrates[43,44]. In this paper, we focus on the local electronic properties of the wrinkles of the twisted graphene bilayer.

**STM and STS measurements.** The STM system was an ultrahigh vacuum four-probe SPM from UNISOKU. All STM and STS measurements were performed at liquid-nitrogen temperature and the images were taken in a constant-current scanning mode. The STM tips were obtained by chemical etching from a wire of Pt(80%) Ir(20%) alloys. Lateral dimensions observed in the STM images were calibrated using a standard graphene lattice. The STS spectrum, i.e., the dI/dV-V curve, was carried out with a standard lock-in technique using a 957 Hz a.c. modulation of the bias voltage.

**Tight binding calculations.** All the calculations were based on tight binding theories. The interlayer Hamiltonian of twisted graphene bilayer reads

$$H_\perp = \sum_{i,\alpha,\beta} t_\perp \left(\delta^{\beta'\alpha}(r_i)\right) c_\alpha^\dagger(r_i) c_{\beta'}\left(r_i + \delta^{\beta'\alpha}(r_i)\right) + H.c..$$

The strain and curvature can change the Slater-Koster overlap integral and modify the hopping parameters, which obey the following two formulas

$$t(l) = t_0 e^{-3.37(l/a - 1)}.$$

$$\tilde{t}_{ij} = \frac{u_{ij}^2}{d_{ij}^4}\left[\left(V_{pp\sigma} - V_{pp\pi}\right)\left(\mathbf{n}_i \cdot \mathbf{d}_{ij}\right)\left(\mathbf{n}_j \cdot \mathbf{d}_{ij}\right) + V_{pp\pi} d_{ij}^2 \mathbf{n}_i \cdot \mathbf{n}_j\right].$$

Based on the revised hopping parameters, the Hamiltonian matrix in the four-component spinor representation is obtained. The band structure of the corrugated twisted bilayer graphene can be calculated through diagonalizing the 4×4 Hamiltonian matrix.

**Acknowledgements**

We acknowledge helpful discussions with Q. Niu and J. Feng. This work was supported by the National Natural Science Foundation of China (Grant Nos. 11004010, 10804010, 10974019, 21073003, 51172029 and 91121012), the Fundamental Research Funds for the

Central Universities, and the Ministry of Science and Technology of China (Grants Nos. 2011CB921903, 2012CB921404, 2013CB921701).

**Author contributions**

W.Y. and L.M. performed the STM experiments of the graphene on Rh foil. W.Y.H. and Z.D.C. performed the theoretical calculations. M.X.L.,Y.F.Z., and Z.F.L. fabricated the samples. L.H. conceived and provided advice on the experiment, analysis, and theoretical calculation. All authors participated in the data discussion and writing of the manuscript.

**Additional information**

The authors declare no competing financial interests.Supplementary Information accompanies this paper at…. Correspondence and requests for materials should be addressed to L. H.

# Supplementary Information

**I.  Supplementary experimental data and data analysis**

In this section we present further experimental data and discussion demonstrating the strain-induced evolution of the electronic band structures of the twisted graphene bilayer and addressing other issues.

Figure S1 shows a typical STM image of the monolayer graphene grown on a Rh (111) surface. The lattice mismatch between graphene (0.246 nm) and Rh(111) (0.269 nm) leads to hexagonal moiré superstructures with the expected periodicity ~ 2.9 nm resulting from a 12C/11Rh coincidence lattice. The characteristic of the Moiré superstructures is distinct from that shown in Fig. 1.

Figure S2 shows the experimental values of $\Delta E_{VHS}$ obtained from the curves 1-7 of Figure 1. It indicates that the $\Delta E_{VHS}$ decreases when the experimental position approaches the strained wrinkle. We demonstrated that the decrease of $\Delta E_{vhs}$ originates partly from the strain-induced lattice deformations. The strained structure may enhance the interlayer coupling, which could also result in the decrease of $\Delta E_{vhs}$.

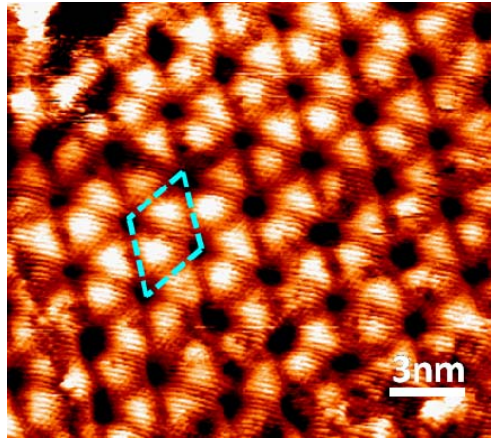

Figure S1. A typical STM image of monolayer graphene grown on Rh (111) surface. It shows moiré pattern with a period of 2.9 nm.

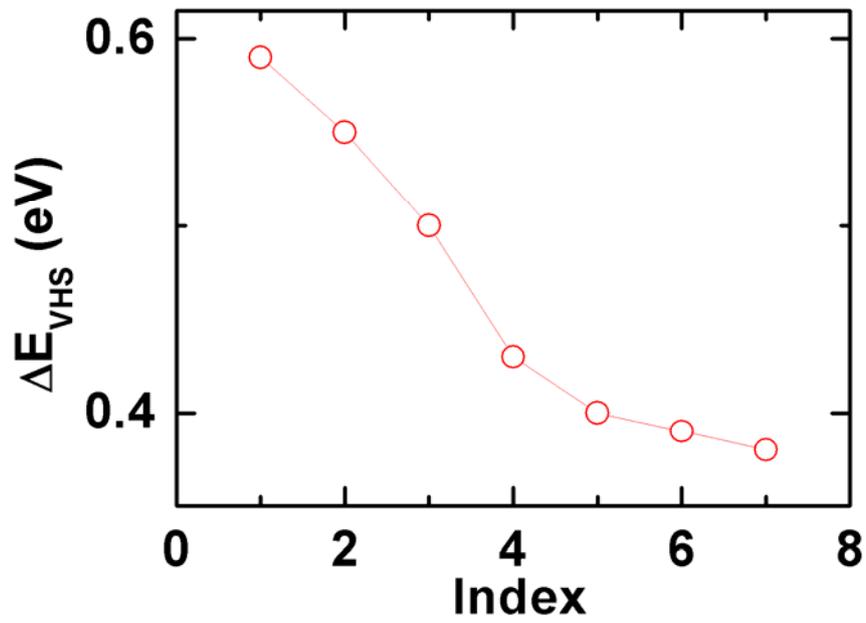

Figure S2. The open circles are the experimental values of $\Delta E_{VHS}$ obtained from curves 1-7 of Figure 1.

Figure S3 shows a STM topography of the strained graphene wrinkle. The strained structure is quite rough along the wrinkle, which may generate local nonuniform strains and consequently result in the large local pseudomagnetic fields. Our experiments also confirm that the pseudomagnetic field is not uniform along the wrinkle, as shown in Figure S4. To further explore the LDOS of the strained wrinkle, we carried out more STS measurements at different positions along the wrinkle. All the STS curves (recorded randomly along the wrinkle) can be divided into two groups showing distinct characteristics. Fig. S4 shows five typical curves of them. For type (I) curves, which were observed more frequently in our experiment, the spectra show LLs following the progression of LLs of the massive Dirac Fermions $E_N = \pm \{\hbar\omega_c[N(N-1)]^{1/2} + E_g/2\}$, $N = 0,1,2,.....$ The pseudomagnetic field and energy gap deduced from these curves varies at different positions along the wrinkle. Usually, the obtained pseudomagnetic field ranges from 90 to 130 T. For type (II) curves, the spectra show LLs at positive bias and a gap with zero tunneling DOS around the Fermi level, but only show weak peaks with much smaller distance in energy at negative bias. This apparent asymmetry of the tunneling spectra is completely reproducible in our experiment. In the "nanobubbles" formed by strained graphene monolayer, the authors also observed asymmetry of the tunneling spectra: they observed clear LLs at positive bias, but not at negative bias.[1] The asymmetry was mainly attributed to the electron-hole asymmetry originating from next-nearest-neighbor hopping. In the strained graphene structures, such large electron-hole asymmetry is expected to observe because of the lattice deformation, which enhances the next-nearest-neighbour hopping.

For graphene wrinkles with much smaller curvature and smooth surface, we can obtain the atomic-resolution STM image, as shown in Figure S5. However, these wrinkles do not show pseudo-Landau quantizations because of the lacking of the non-uniform strains. Another grpahene wrinkle with obvious moiré pattern on the top region of the wrinkle is shown in Figure S6.

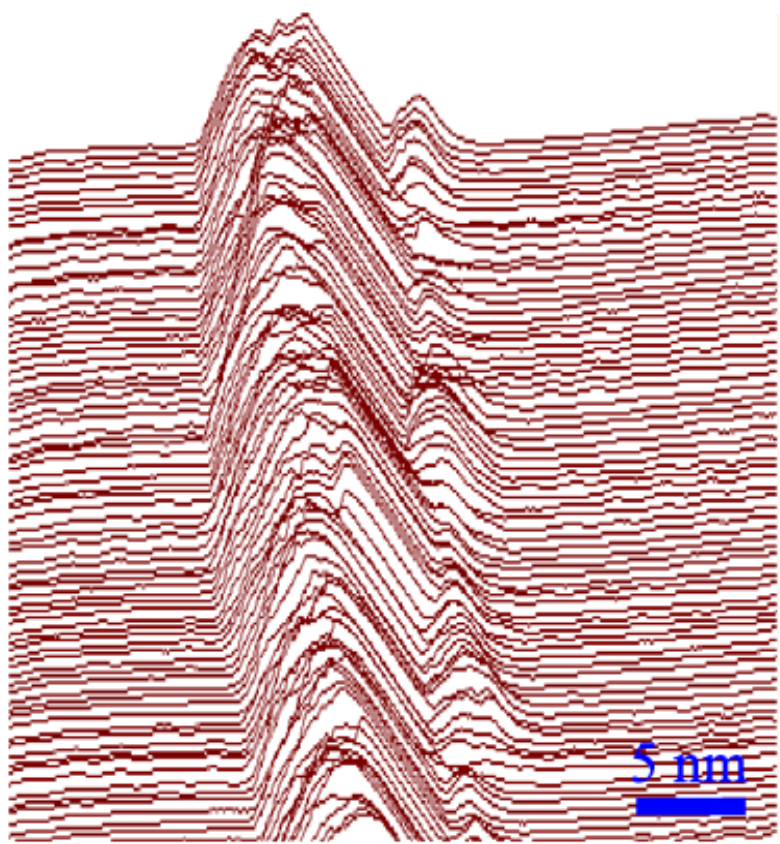

Figure S3. A STM topography of the strained graphene wrinkle.

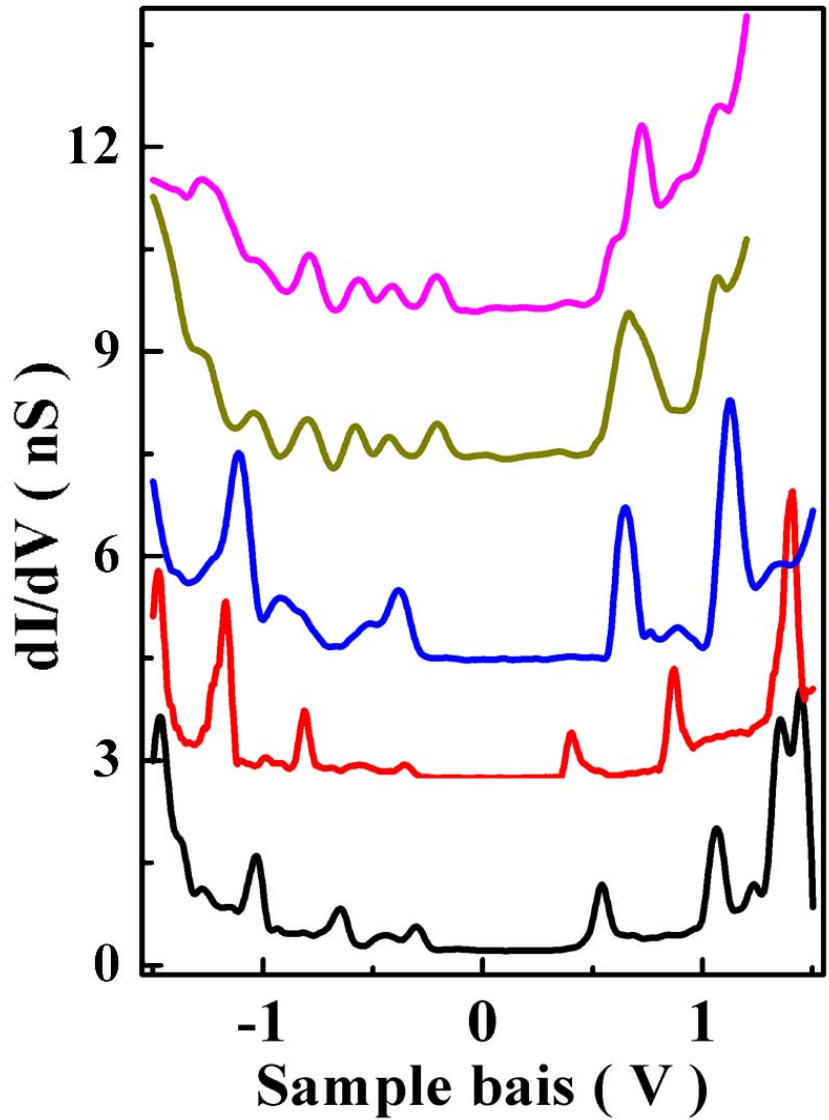

Fig. S4. Five typical dI/dV-V curves taken along the strained wrinkle. All the spectra can be divided into two groups with distinct characteristics: (I) the three curves in the lower panel show LLs of massive Dirac fermions with a bandgap; (II) the two curves in the upper panel show LLs at positive bias, but only show weak peaks with much smaller energy interval at negative bias.

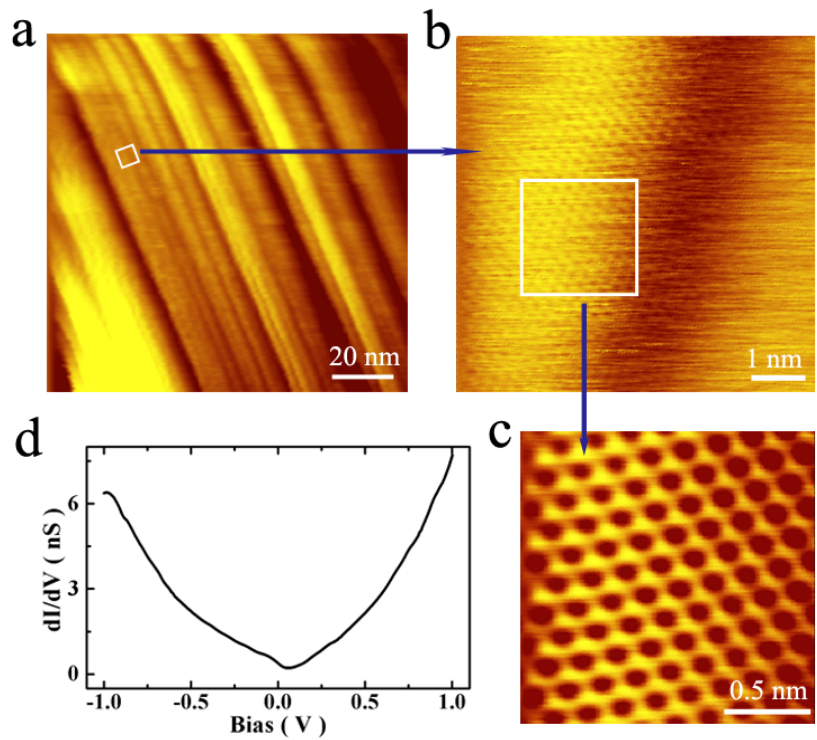

Figure S5. **a** Large-scale STM image showing the typical surface morphology of graphene growth on Rh foils under atmospheric growth conditions. The stripped protrusions along the terraced steps are the graphene wrinkles, usually presenting a height of 1-2 nanometers and a lateral width of several tenths of nanometers. **b** Zoom-in topography of the white frame in (**a**). **c** Zoom-in topography of the white frame in (**b**) shows atomic-resolution image. **d** A typical STS obtained along the graphene wrinkles. No pseudo-Landau quantization is observed due to the small curvature (strain) and smooth surface (no non-uniform strain).

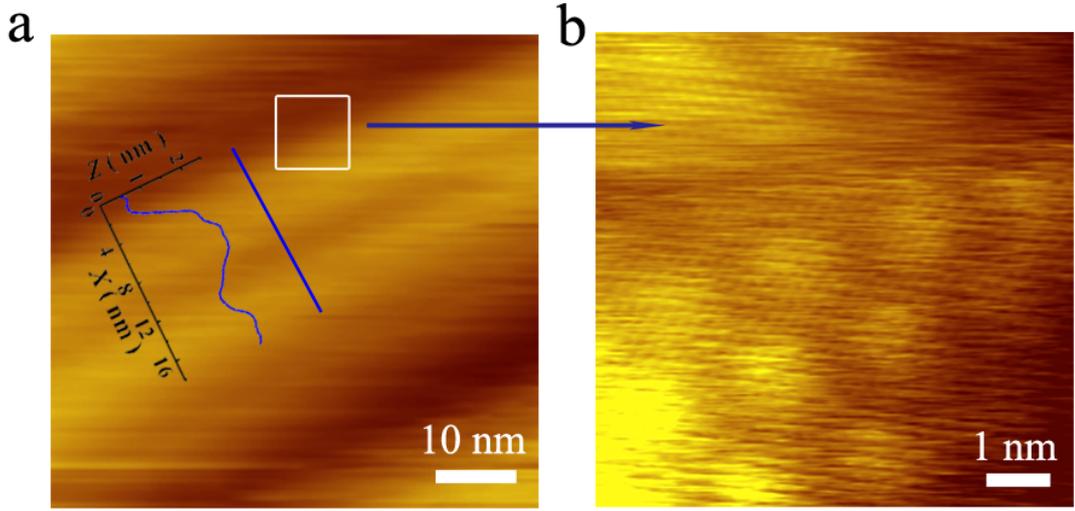

Figure S6. **a** Large scale STM of graphene wrinkle on Rh foil ($V_{sample}$ = -204 mV and I = 0.23nA). The inset shows a line profile along the blue straight line. **b** Zoom-in topography of the write frame in (a) shows obvious moiré pattern with period of 2.1 nm on the top region of the wrinkle ($V_{sample}$=-203mV and I=0.27nA).

## II. Theory

(a) Tight binding model for the twisted graphene bilayer:

The Hamiltonian for the bilayer with a twist has the form $H = H_1 + H_2 + H_\perp$, where $H_1$ and $H_2$ are the Hamiltonians for each layer, $H_\perp$ is the interaction Hamiltonian between the two layers following the model in Refs. 4-6. The Hamiltonians of $H_1$, $H_2$, and $H_\perp$ can be expressed as

$$H_1 = -t_0 \sum_l a_1^\dagger(\mathbf{r}_i)[b_1(\mathbf{r}_i + \boldsymbol{\delta}_1) + b_1(\mathbf{r}_i + \boldsymbol{\delta}_2) + b_1(\mathbf{r}_i + \boldsymbol{\delta}_3)] + H.C.$$

$$H_2 = -t_0 \sum_j a_2^\dagger(\mathbf{r_j})[b_2(\mathbf{r_j}+\boldsymbol{\delta_1})+b_2(\mathbf{r_j}+\boldsymbol{\delta_2})+b_2(\mathbf{r_j}+\boldsymbol{\delta_3})+H.C.]$$

$$H_\perp = -t_\perp \sum_i a_1^\dagger(\mathbf{r_i})a_2(\mathbf{r_i})+H.C.$$

(1).

Here the index m (1 or 2) represents the layer and $a_m(r)(b_m(r))$ is the annihilation term on sublattice A(B), at site **r**. $\boldsymbol{\delta_\alpha}$ connects the site **r** to its neighbors, where **α**=1,2,3, $\boldsymbol{\delta_1}$=(0, 1)a, $\boldsymbol{\delta_2}$=(-√3/2, -1/2)a, $\boldsymbol{\delta_3}$=(√3/2, -1/2 )a, and a($\approx$1.42Å) is the nearest carbon-carbon distance. The intra- and inter-plane hopping energies are $t_0$ and $t_\perp$, respectively. In our model, we use the values $t_0$= 3 eV and $t_\perp$= 0.4 eV. Two saddle points (VHSs) form at $k$ = 0 between the two Dirac cones **K** and **K$_\theta$** with a separation of $\Delta K$ =2$K$sin($\theta$/2). The low-energy VHSs, as shown in Fig. S7, contribute to two pronounced peaks flanking zero-bias in a typical tunneling spectrum obtained in our experiment. The detail of the calculation could be found in Ref. 4-7.

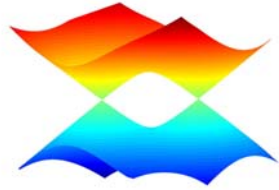

Figure S7. Electronic band structure of twisted bilayer graphene with a finite interlayer coupling calculated with the four-band model. Two saddle points (VHSs) form at $k$ = 0 between the two Dirac cones, $K$ and $K_\theta$, with a separation of $\Delta K$ =2$K$sin($\theta$/2).

(b) Tight binding model for the emergence of superlattice Dirac points in the twisted graphene bilayer:

The moiré pattern, both induced between the top layer graphene and the substrate (or the under layer graphene), could provide a weak periodic potential. Here we consider the possible emergence of the superlattice Dirac points in the twisted graphene bilayer induced by the graphene-on-graphene moiré. The bilayer graphene with a twist angle that satisfies the condition for commensurate periodic structure leads to the moiré patterns. With considering that the graphene-on-graphene moiré patterns provide an external periodic potential, we can use the model developed for graphene monolayer in a periodic potential[8].

To further simplify our calculation, we assume that the moiré pattern generates a weak muffin-tin type of periodic potential on the twisted graphene bilayer with the potential value $\Delta U$ in a triangular array of disks of diameter d and zero outside of the disks. The spatial period of the superlattice is L, which equals to the period of the moiré pattern (here d < L). Then the total Hamiltonian of a twisted graphene bilayer in a periodic potential can be written as:

$$H = \left[ H_1 + \Delta U \sum_\alpha \cos(\vec{G}_\alpha \cdot \vec{x}_1) \begin{pmatrix} I & 0 \\ 0 & 0 \end{pmatrix} \right] + \left[ H_2 + \Delta U \sum_\alpha \cos(\vec{G}_\alpha \cdot \vec{x}_2) \begin{pmatrix} 0 & 0 \\ 0 & I \end{pmatrix} \right] + H_\perp$$

where $I$ is an identity matrix, and $G_\alpha$ is the potential's reciprocal lattice. By using a unitary transformation $H^*=M^\dagger H M$, where $M = \dfrac{\sqrt{2}}{2}\begin{pmatrix} e^{-i\eta(x,y)/2} & e^{-i\eta(x,y)/2} \\ e^{-i\eta(x,y)/2} & e^{i\eta(x,y)/2} \end{pmatrix}$ and

$\eta(x,y) = 2\int_0^x \int_0^y V(x',y')dx'dy'/\hbar v_0$, then the energy eigenvalues can be worked out through diagonalizing the Hamiltonian matrix[9]. The superlattice Dirac points of the twisted graphene bilayer generated by the graphene-on-graphene moiré is shown in Fig. S8.

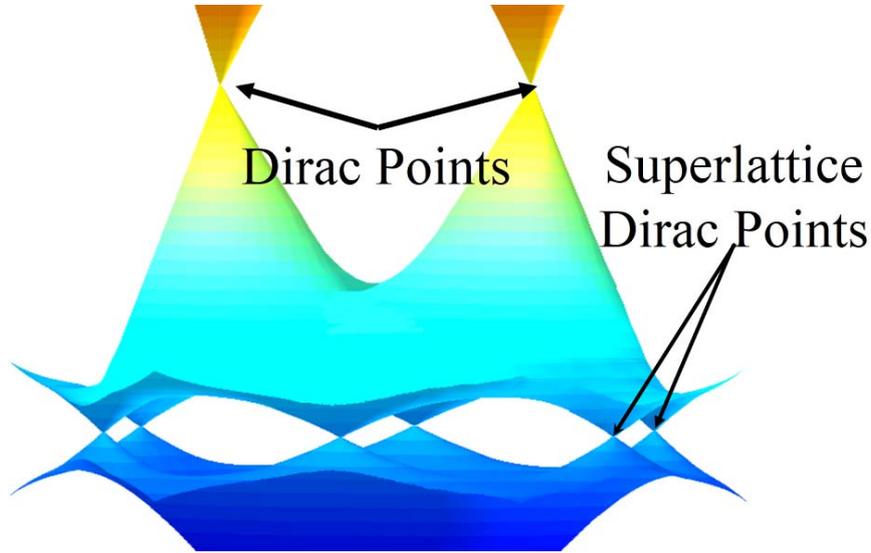

Figure S8. Energy of charge carriers in a twisted graphene bilayer showing the original Dirac points and the emergence of new superlattice Dirac points.

(c) Tight binding model for the corrugated twisted graphene bilayer:

The Hamiltonian $H = H_1 + H_2 + H_\perp$, can be simplified into a 4×4 matrix after introducing four-component spinors

$$\Psi_k^\dagger = \left(a_1^\dagger, b_1^\dagger, a_2^\dagger, b_2^\dagger\right) \quad (2)$$

In this four-component spinors representation, the Hamiltonian reads

$$H(\mathbf{k}) = \begin{pmatrix} H_{11} & H_{12} \\ H_{21} & H_{22} \end{pmatrix} \quad (3)$$

with

$$H_{11} = H_{22} = \begin{pmatrix} 0 & \zeta(\mathbf{k}) \\ \zeta(\mathbf{k})^* & 0 \end{pmatrix},$$

$$H_{12} = \begin{pmatrix} t_\perp & 0 \\ 0 & 0 \end{pmatrix},$$

and

$$\zeta(\mathbf{k}) = -\sum_{i=1}^{3} t_0 e^{i\mathbf{k}\cdot\boldsymbol{\delta}_i}.$$

Here, $H_{11}$ and $H_{22}$ are the intra-layer Hamiltonian, while $H_{12}$ and $H_{21}$ describe the inter-layer coupling between the two layers. The strain in graphene would change the nearest carbon-carbon distance, while a small twisted angle between two layers is expected to have an effect upon the inter-layer coupling.

In the flat twisted graphene bilayer, only the in plane strain was taking into account since that the curvature is very small. On the assumption that the twisted bilayer graphene is uniformly compressed along the armchair direction of layer 1 (this consists with our experimental result), then the strain tensor in both layers can be written as

$$\varepsilon_1 = -\varepsilon \begin{pmatrix} -\sigma & 0 \\ 0 & 1 \end{pmatrix}. \quad (4)$$

In the presence of the strain, the lattice of graphene would be deformed. In layer 1, the deformed bonds are given by

$$\delta_{11}^{T} = (I+\varepsilon)\delta_{1}^{T},$$

$$\delta_{12}^{T} = (I+\varepsilon)\delta_{2}^{T},$$

$$\delta_{13}^{T} = (I+\varepsilon)\delta_{3}^{T},$$

and in layer 2, the deformed bonds change to

$$\delta_{21}^{T} = (I+\varepsilon)\begin{pmatrix} \cos\theta & -\sin\theta \\ \sin\theta & \cos\theta \end{pmatrix}\delta_{1}^{T},$$

$$\delta_{22}^{T} = (I+\varepsilon)\begin{pmatrix} \cos\theta & -\sin\theta \\ \sin\theta & \cos\theta \end{pmatrix}\delta_{2}^{T},$$

$$\delta_{23}^{T} = (I+\varepsilon)\begin{pmatrix} \cos\theta & -\sin\theta \\ \sin\theta & \cos\theta \end{pmatrix}\delta_{3}^{T},$$

where I is the unit matrix. Due to the variation in the length of the carbon-carbon bond, the hopping parameters would be no longer isotropic. The new hopping parameters can be obtained according to[10]

$$t(l) = t_0 e^{-3.37(l/a - 1)}, \quad (5)$$

Where $l$ is the bond length and $a$ is the initial length of carbon-carbon bond. Therefore, in layer 1, the new hopping parameters are

$$t_{11} = t_0 e^{-3.37(|\delta_{11}|/a - 1)},$$

$$t_{12} = t_0 e^{-3.37(|\delta_{12}|/a - 1)},$$

$$t_{13} = t_0 e^{-3.37\left(|\boldsymbol{\delta}_{13}|/a - 1\right)},$$

and in layer 2

$$t_{21} = t_0 e^{-3.37\left(|\boldsymbol{\delta}_{21}|/a - 1\right)},$$

$$t_{22} = t_0 e^{-3.37\left(|\boldsymbol{\delta}_{22}|/a - 1\right)},$$

$$t_{23} = t_0 e^{-3.37\left(|\boldsymbol{\delta}_{23}|/a - 1\right)}.$$

Hence, by substituting new hopping parameters and neighbor vectors in unstrained Bernal bilayer graphene, we can derive the new $H_{11}$ and $H_{22}$, which are mainly responsible for the change in the intra-layer hopping parameters.

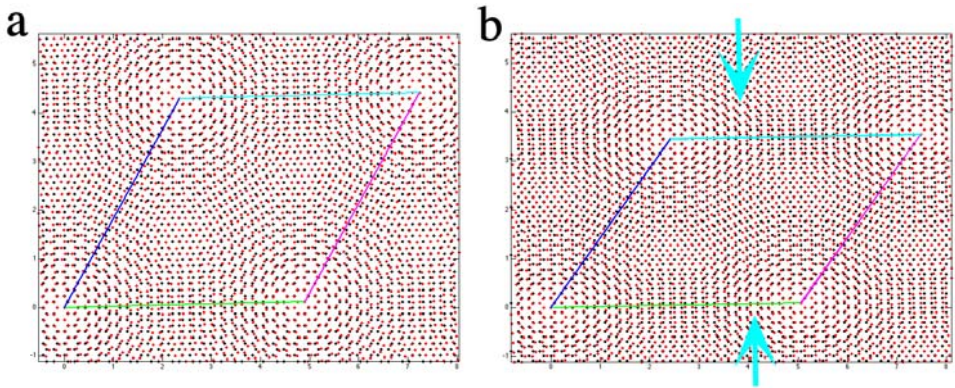

Figure S9. Structural model of a twisted graphene bilayer in a tension strain. The twisted graphene bilayer is uniformly compressed along a prescribed direction. (a) Standard structure of a twised bilayer graphene. (b) Structure of the twisted bilayer graphene in a compressed strain.

For the twisted bilayer graphene, after adding a uniform compression along the armchair direction (layer 1), the whole moire pattern is also squeezed, which is shown in Fig. S9. It shows that the moire pattern is retained and the superlattice basis vectors are also compressed. As a result, we can follow the methods in Refs. 6, 10-12 to deal with the interlayer hopping parameters. The interlayer Hamiltonian reads

$$H_\perp = \sum_{i,\alpha,\beta} t_\perp\left(\delta^{\beta'\alpha}(r_i)\right) c_\alpha^\dagger(r_i) c_{\beta'}\left(r_i + \delta^{\beta'\alpha}(r_i)\right) + H.c. \quad (6)$$

For the compressed commensurate structure, the Fourier transformation in Ref. 6 still holds. The reciprocal lattice vectors are changed due to the compressed strain. Since the compressed commensurate structure remains its original symmetry, the Fourier amplitudes of $\tilde{t}_\perp^{AB}, \tilde{t}_\perp^{AA}, \tilde{t}_\perp^{BB}$ can be expressed in terms of $\tilde{t}_\perp^{BA}$. The new reciprocal lattice vectors can be obtained through applying the strain tensor as before. Then following the model in Refs. 6, 10-12, and setting our interlayer hopping $t_\perp$=0.4 eV, we can derive the same result as shown in Table I of Ref. 11. Then in our four-component spinors representation, these three main Fourier components leads to three types of interlayer hopping terms[12,13],

$$H_\perp^0 = \tilde{t}_\perp \begin{pmatrix} 1 & 1 \\ 1 & 1 \end{pmatrix}, \quad H_\perp^\pm = \tilde{t}_\perp \begin{pmatrix} e^{\mp i\phi} & 1 \\ e^{\pm i\phi} & e^{\mp i\phi} \end{pmatrix}, \quad (7)$$

where $\phi = 2\pi/3$, and $\tilde{t}_\perp = 0.4 t_\perp$.[6,13]

Based on the revised $H_{11}$, $H_{12}$, $H_{21}$, $H_{22}$, the Hamiltonian matrix in the four-component spinor representation is obtained. The band structure of the corrugated twisted bilayer graphene can be calculated through diagonalizing the 4×4 Hamiltonian matrix, as shown in the main text.

At the top of the wrinkle, the curvature radius, ~ 1.41 nm, is obtained through a Gaussian curve fitting to the section line of the wrinkle. As the bond length of carbon atoms is much smaller than the curvature radius, the curvature will rehybridize the π and σ orbitals of graphene and cause a change in Slater-Koster overlap integral[14-16]. According to ref. 16, in curved space $t_{ij}$ is replaced by

$$\tilde{t}_{ij} = \frac{u_{ij}^2}{d_{ij}^4}\left[\left(V_{pp\sigma} - V_{pp\pi}\right)\left(\mathbf{n}_i \cdot \mathbf{d}_{ij}\right)\left(\mathbf{n}_j \cdot \mathbf{d}_{ij}\right) + V_{pp\pi}d_{ij}^2 \mathbf{n}_i \cdot \mathbf{n}_j\right], \quad (7)$$

Here, $\mathbf{u}_{ij}$ is the vector connecting atoms i and j in the undeformed lattice while $\mathbf{d}_{ij}$ is the corresponding vector after deformation. At the top of the wrinkle, the strain still exists, so $V_{pp\pi}$ obeys equation (5) and $V_{pp\sigma} = 1.7 V_{pp\pi}$[15]. Dealing these geometrical quantities in the circle of curvature and substituting them into equation (7), we can obtain new hopping parameters induced by the joint effect of strain and curvature.